\documentclass[11pt]{article}
\usepackage{fontspec}
\usepackage{amsmath,amssymb,amsthm}
\usepackage{fancyvrb}
\usepackage{booktabs}
\usepackage{stmaryrd}
\usepackage[margin=2.7cm]{geometry}
\usepackage{xcolor}
\usepackage[colorlinks,linkcolor=black,citecolor=black,urlcolor=blue!60!black]{hyperref}

\DefineVerbatimEnvironment{agda}{Verbatim}{xleftmargin=1.5em,fontsize=\small}
\newcommand{\ag}[1]{\texttt{\detokenize{#1}}}
\newcommand{\repourl}{https://github.com/txa/sst-new}
\newcommand{\agfile}[1]{\href{\repourl/blob/main/#1}{\ag{#1}}}
\newcommand{\agdir}[1]{\href{\repourl/tree/main/#1}{\ag{#1/}}}

\newtheorem*{remark}{Remark}

\title{Any colour you like, as long as it's UIP\\[0.5ex]
\large Semi-simplicial types from weakenings and thinnings}
\author{Thorsten Altenkirch\\[0.5ex]
\small University of Nottingham, United Kingdom\\
\small\href{mailto:Thorsten.Altenkirch@nottingham.ac.uk}{\ag{Thorsten.Altenkirch@nottingham.ac.uk}}}
\date{October 2026}

\begin{document}
\maketitle

\begin{abstract}
We explain an attempt to define semi-simplicial types (SST) in homotopy type
theory by first defining a small type theory whose contexts are exactly the
data of semi-simplicial types, constructing $\mathrm{SST}\,n$ uniformly in that
syntax, and then interpreting the syntax in the universe. The syntax can be
defined without any truncation and without K, and the construction of
$\mathrm{SST}\,n$ goes through. The naive interpretation does not: it
transports along the semantic images of equations that hold only
propositionally, and making these transports agree needs uniqueness of
identity proofs (UIP) for the semantic types. Moving the problem into the
syntax, by making composition a constructor of a higher inductive-inductive
type (HIIT), does not remove it either: it
reappears as a coherence problem in the syntax. Fording the HIIT (storing
equations instead of computed indices), with conditional rules, makes the
syntax provably a set. This is UIP for the
syntax, obtained by encode-decode, and it does not by itself restrict the
semantics; but the naive interpretation of the conditional rules again needs
UIP for the semantic types. A semantics into groupoids that avoids UIP will be
discussed in a separate note. The Agda development accompanying this note is in the repository
\href{\repourl}{\ag{github.com/txa/sst-new}}.
\end{abstract}

\section{Why semi-simplicial types?}

Homotopy type theory describes $\infty$-groupoids: every type comes with its
identity types, and J supplies all their coherences for free. It does not
(yet) describe higher \emph{categories}, where morphisms need not be
invertible. The standard way to present an $(\infty,1)$-category is by a
(semi-)simplicial object of types satisfying Segal and completeness
conditions, and Capriotti and Kraus showed that this works in a type theory
where semi-simplicial types are available~\cite{capriotti-kraus}. So if we
could define the type of semi-simplicial types in HoTT, we could represent
higher categories internally, including the $(\infty,1)$-category of types
itself.

A closely related goal is for HoTT to ``eat itself'': to define the syntax of
HoTT inside HoTT and interpret it in the universe. Shulman's account of this
problem~\cite{shulman-eat} singles out semi-simplicial types as one of the
most important open problems on the way, since they are the simplest
structure that needs an infinite tower of coherences. (The title echoes
Chapman's \emph{Type theory should eat itself}~\cite{chapman}.)

The question of defining semi-simplicial types in HoTT was raised during the
special year on Univalent Foundations at the IAS in 2012--13, by Voevodsky
and others~\cite{ufias-sst}. Voevodsky's proposal of a type system with two
identity types, one strict and one homotopical~\cite{voevodsky-hts}, makes
it possible, and this idea was developed into two-level type
theory~\cite{ack-strict,acks-2ltt}, in which semi-simplicial types can be
defined. Herbelin gave a construction~\cite{herbelin} which, as he later
pointed out, relies on uniqueness of identity proofs. Kolomatskaia and
Shulman define them in displayed type theory~\cite{dtt}. In book HoTT the
problem remains open.

\section{Semi-simplicial types up to triangles}

A semi-simplicial type up to level 2 consists of
\begin{align*}
X_0 &: \mathcal U\\
X_1 &: X_0 \to X_0 \to \mathcal U\\
X_2 &: (x_0\,x_1\,x_2 : X_0) \to X_1\,x_0\,x_1 \to X_1\,x_0\,x_2 \to X_1\,x_1\,x_2 \to \mathcal U
\end{align*}
Points, edges between points, and triangles filling a boundary of three
points and three edges. The problem is to define, for a variable $n:\mathbb N$,
the type of such data up to level $n$. For each numeral $n$ we can write it
down, but nobody knows how to do it uniformly in $n$ in book HoTT.

Two observations about the example guide everything that follows.
\begin{enumerate}
\item Each $X_k$ is a family over a \emph{telescope} of arguments, and each
  entry of that telescope is an earlier family applied to a \emph{selection}
  of earlier entries. The edge $x_{02} : X_1\,x_0\,x_2$ selects $x_0$ and
  $x_2$ from the three points.
\item The selections have to be compatible. The arguments of $X_1$ are two
  points, so the selection that picks out $x_{02}$ must also pick out the two
  points it connects. Selections compose, and at level 2 this composition is
  used for the first time in a non-trivial way: the boundary of a triangle
  consists of edges, and the boundary of an edge consists of points.
\end{enumerate}
Level 2 is therefore the smallest example in which everything below happens.

\section{A type theory with only weakenings and thinnings}

We describe the data above by a small type theory with five sorts:
\begin{center}
\begin{tabular}{lll}
\toprule
\ag{Con} & contexts & sequences of type families $X_0, X_1, \dots$\\
\ag{Ty Γ} & types & telescopes (domains of families)\\
\ag{Var Γ A} & variables & a family in $\Gamma$ over the telescope $A$\\
\ag{Wk Δ Γ} & weakenings & forgetting the last families of a context\\
\ag{Thin A B} & thinnings & projections from telescope $A$ to $B$\\
\bottomrule
\end{tabular}
\end{center}
There are two directions of movement:
\begin{itemize}
\item \emph{vertically}, a weakening goes from a longer to a shorter context:
  one more family is in scope. We only need one step,
  \ag{p : Wk (Γ ▷ A) Γ}, and no general substitution;
\item \emph{horizontally}, within a fixed context, a thinning goes from a
  longer to a shorter telescope by dropping some entries and keeping others.
\end{itemize}
The constructors are
\begin{agda}
•    : Con
_▷_  : (Γ : Con) → Ty Γ → Con          -- add a family over a telescope
ε    : Ty Γ                            -- empty telescope
ext  : (A : Ty Γ) → Var Γ B → Thin A B → Ty Γ
                                       -- A extended by an entry  x (δ a)
zero : Var (Γ ▷ A) (wkTy A p)
suc  : Var Γ A → Var (Γ ▷ B) (wkTy A p)
tε   : Thin ε ε
t-d  : Thin A B → Thin (ext A x δ) B   -- drop the last entry
t-k  : Thin A B → Thin (ext A x ?) (ext B x δ)   -- keep it
\end{agda}
The variables \ag{zero} and \ag{suc} need weakening: \ag{zero} refers to the
last family of \ag{Γ ▷ A}, whose telescope \ag{A} lives in the shorter context
\ag{Γ}, so its type is \ag{A} moved along \ag{p}. Weakening is not a
constructor but an operation, defined by recursion on the syntax, which
pushes the weakening through types, variables and thinnings:
\begin{agda}
wkTy   : Ty Γ → Wk Δ Γ → Ty Δ
wkVar  : Var Γ A → (w : Wk Δ Γ) → Var Δ (wkTy A w)
wkThin : Thin A B → (w : Wk Δ Γ) → Thin (wkTy A w) (wkTy B w)

wkTy ε           w = ε
wkTy (ext A x δ) w = ext (wkTy A w) (wkVar x w) (wkThin δ w)
wkVar x p          = suc x
wkThin tε      w   = tε
wkThin (t-d σ) w   = t-d (wkThin σ w)
wkThin (t-k σ) w   = t-k (wkThin σ w)
\end{agda}
Weakening a telescope only renames the families it refers to (each $x$
becomes \ag{suc x}); the shape of the telescope and of its thinnings is
unchanged. For example the edges $E_{ij}$ above have target
\ag{wkTy A₁ p}: the two-point telescope $A_1$ moved into the context where
$X_1$ is in scope.
In the intended interpretation a context is a nested $\Sigma$-type of
families, a telescope $A$ is a type depending on the context, \ag{ext A x δ}
is $\Sigma\,(a : A)\,.\,x\,(\delta\,a)$, and a thinning is a projection.

In this language the example reads:
\begin{agda}
Γ₀ = • ▷ ε                                        -- X₀
A₁ = ext (ext ε zero tε) zero (t-d tε)            -- two points
Γ₁ = Γ₀ ▷ A₁                                      -- X₁
V₃ = three points (variable suc zero)
E₀₁ E₀₂ E₁₂ : Thin V₃ (wkTy A₁ p)                 -- the three edges
A₂ = ext (ext (ext V₃ zero E₀₁) zero (t-d E₀₂)) zero (t-d (t-d E₁₂))
\end{agda}
The thinnings $E_{ij}$ select the endpoints of the three edges, for example
\ag{E₁₂ = t-k (t-k (t-d tε))} drops $x_0$ and keeps $x_1, x_2$.

\paragraph{What is the \ag{?}?}
If we keep the last entry $x\,(\delta\,b)$ of \ag{ext B x δ} while thinning
the rest along $\sigma : \ag{Thin}\,A\,B$, the entry of the source must be
$x$ applied to the composite selection: the index is $\delta\circ\sigma$.
\begin{agda}
t-k : (σ : Thin A B) → Thin (ext A x (δ ∘ σ)) (ext B x δ)
\end{agda}
This is where composition enters. In \ag{E₁₂}, keeping $x_2$ requires
$x_2$'s stored selection in $V_3$ (drop $x_0$ and $x_1$) to be the composite
of the selection in $A_1$ (drop the first point) with the thinning that drops
$x_0$.

\section{The syntax: induction-recursion}\label{sec:ir}

We want the syntax to be an ordinary inductive definition, \emph{not} a
quotient. A quotient inductive-inductive type is a set, and its recursor only
eliminates into sets, whereas we want to interpret into the universe, which is
not a set. So the syntax must be untruncated: equality of syntax is the
equality of an inductive type, and it is a theorem (not an axiom) that it is a
set.

The type of \ag{t-k} mentions \ag{δ ∘ σ}, and the types of \ag{zero} and
\ag{suc} mention \ag{wkTy}. So composition and weakening have to be defined
\emph{simultaneously} with the syntax: the syntax is an inductive-inductive-recursive
definition, in which functions defined by recursion on the types being
defined occur in the indices of their constructors. (In the code we package
the sorts as one family \ag{El : U → Set} indexed by codes; this is only a
convenient way to write the mutual definition in Agda, not a reduction to
something simpler.)

We should be honest about the status of this. Agda accepts such definitions,
but they are a feature of Agda whose semantics is not well understood. There
are established accounts of induction-recursion~\cite{dybjer-ir,dybjer-setzer}
and of induction-induction~\cite{forsberg-setzer}, but none that we know of
covers definitions like ours, where recursively defined functions, and in the
later versions also proofs about them, appear in constructor indices. We use
it anyway, as a powerful piece of ``magic'' that lets us state and check the
constructions. The point of the rest of this note is that even with this
magic we still run into an obstruction.

\paragraph{First attempt: index by the composite.}
With \ag{t-k} as above, define composition by recursion on the right
argument:
\begin{agda}
δ      ∘ tε    = δ
δ      ∘ t-d σ = t-d (δ ∘ σ)
t-d δ  ∘ t-k σ = t-d (δ ∘ σ)
t-k δ  ∘ t-k σ = t-k (δ ∘ σ)          -- ill-typed!
\end{agda}
The left side has type \ag{Thin (ext A x ((δC ∘ δ) ∘ σ)) _}, the right side
\ag{Thin (ext A x (δC ∘ (δ ∘ σ))) _}. They agree only up to associativity, a
theorem we prove by induction, so we would have to write
\ag{subst _ ∘-assoc (t-k (δ ∘ σ))}. The same happens for the identity
(\ag{t-k id} has index \ag{δ ∘ id}, needing \ag{∘-id}) and for weakening a
thinning (needing \ag{wkThin (δ ∘ σ) ≡ wkThin δ ∘ wkThin σ}). A \ag{subst}
along a propositional equation in a definition blocks computation in
everything that uses it. In addition, pattern matching on \ag{t-k} forces Agda
to unify an index with \ag{δ ∘ σ}, a defined function (``green slime'', in McBride's terminology).

\paragraph{Second attempt: Fording.}
Following McBride, we store the equation instead of computing the
index:\footnote{The name alludes to Henry Ford's slogan that a customer can
have a car painted any colour, so long as it is black. Instead of restricting
the index of \ag{t-k} to be \ag{δ ∘ σ}, we allow any index \ag{δ'}, so long as
it is equal to \ag{δ ∘ σ}. The name is due to Conor McBride, who speaks of
reformulating an indexed family ``in the Henry Ford tradition''
\cite[Chapter~6]{mcbride-metaprog}; the encoding of indices by equations is also
used in~\cite{levitation}. We have not found an earlier written source;
the phrase may have been used in talks before.}
\begin{agda}
t-k : (σ : Thin A B) → δ ∘ σ ≡ δ' → Thin (ext A x δ') (ext B x δ)
\end{agda}
Now composition is defined without any transport; the equation is built as
data:
\begin{agda}
t-k δ q ∘ t-k σ e = t-k (δ ∘ σ) (∘-k q e)
  where ∘-k : ρ ∘ δ ≡ δB → δB ∘ σ ≡ δA → ρ ∘ (δ ∘ σ) ≡ δA
\end{agda}
and \ag{∘-k} uses \ag{∘-assoc}, which is proved mutually with \ag{_∘_}.
Identity and weakening are handled the same way. Closed terms compute: the
example above is typechecked with all stored equations being \ag{refl}, and
\ag{id ∘ id ≡ id} holds by \ag{refl} at $A_2$.

\paragraph{The new problem: UIP in the syntax.}
To prove the laws (associativity, identity) we need a congruence for
\ag{t-k}:
\begin{agda}
t-k-cong : σ ≡ σ' → t-k σ e ≡ t-k σ' e'
\end{agda}
where \ag{e} and \ag{e'} are two proofs of the same equation
\ag{δ ∘ σ ≡ δ'}. This needs them to be equal, i.e.\ uniqueness of identity
proofs for \ag{Thin}. In Agda this is K (directory \agdir{withK}).

\section{Removing UIP: codes for equality}\label{sec:codes}

The syntax is finitary and first order, so all its sorts have decidable
equality, and by Hedberg's theorem~\cite{hedberg} they are sets. Rather than going through
Hedberg we encode equality of thinnings directly, as a type defined by
recursion (the encode-decode method~\cite{licata-shulman}, \cite[\S 8.1]{hott-book}):
\begin{agda}
Code : Thin A C → Thin A' C' → Set
Code tε          tε          = ⊤
Code (t-d σ)     (t-d τ)     = Code σ τ
Code (t-k σ _ _) (t-k τ _ _) = Code σ τ
Code _           _           = ⊥

t-k : (σ : Thin A B) → CodeV x y → Code (δ ∘ σ) δ'
    → Thin (ext A x δ') (ext B y δ)
\end{agda}
\begin{itemize}
\item Codes are propositions by a plain induction (\ag{Code-prop}): no K, no
  Hedberg, no \ag{Prop}.
\item The kept variable is split into \ag{x} and \ag{y}, related by a code
  \ag{CodeV x y}, so that no index is repeated and pattern matching works
  \ag{--without-K}.
\item Codes can be decoded into equations (\ag{Code-sound}), and the total
  space of variables is a set (\ag{VarΣ-set}).
\end{itemize}
With this, \ag{t-k-cong} holds without K. This achieves the first goal: an
\emph{untruncated} syntax, defined by induction-recursion without K (file
\agfile{TT.agda}), in which stored proofs are unique. That its sorts are sets
follows by the same encode-decode method, using that a reflexive,
proposition-valued relation implying equality makes a type a
set~\cite[Theorem~7.2.2]{hott-book}. For thinnings the ingredients
(\ag{Code-refl}, \ag{Code-prop}, \ag{Code-sound}) are in \agfile{TT.agda};
the full statement for all sorts is formalised for a variant of the syntax in
\agfile{experiments/forded-hiit/NF2.agda} (Section~\ref{sec:fhiit}).

\section{Semi-simplicial types in the syntax}

Following Section~3 of~\cite{ack-strict}, we define by structural
recursion on $k$, simultaneously,
\begin{itemize}
\item the context \ag{ctx k} of levels $X_0,\dots,X_{k-1}$,
\item the $k$-skeleton \ag{SK k n : Ty (ctx k)} of an $(n-1)$-simplex,
\item its face maps \ag{SK→ k f : Thin (SK k n) (SK k m)} for order preserving
  embeddings $f$,
\item their functoriality \ag{α k f g : SK→ k g ∘ SK→ k f ≡ SK→ k (f ⊙ g)}.
\end{itemize}
Level $k+1$ only uses level $k$ weakened once, so all types agree
definitionally. The transport along $\alpha_k$ in the paper becomes the
equation stored in \ag{t-k}, and the coherence of $\alpha$, for which the
paper used K, becomes \ag{t-k-cong}, which is now K-free. The result is
\begin{agda}
SST : ℕ → Con
\end{agda}
defined uniformly in $n$, without K (file \agfile{SSTk.agda}).

\section{The naive interpretation, and the obstruction}\label{sec:interp}

To finish we need the standard interpretation of the syntax in the universe,
defined by recursion on the syntax:
\begin{agda}
⟦ Γ ⟧C  : Set₁                         -- nested Σ of families
⟦ A ⟧T  : ⟦ Γ ⟧C → Set
⟦ x ⟧V  : (γ : ⟦ Γ ⟧C) → ⟦ A ⟧T γ → Set
⟦ σ ⟧Th : ⟦ A ⟧T γ → ⟦ B ⟧T γ           -- projections

⟦ ext A x δ ⟧T γ = Σ (⟦ A ⟧T γ) (λ a → ⟦ x ⟧V γ (⟦ δ ⟧Th a))
\end{agda}
Then $\ag{⟦ SST n ⟧C}$ would be the type of semi-simplicial types up to level
$n$.

\paragraph{Where it breaks.}
Interpreting a kept entry needs the interpretation of the stored equation:
\begin{agda}
⟦ t-k σ e ⟧Th (a , v) = ⟦ σ ⟧Th a , subst ⟦ x ⟧V (keq e a) v
  where keq e a : ⟦ δ' ⟧Th a ≡ ⟦ δ ⟧Th (⟦ σ ⟧Th a)
\end{agda}
and \ag{keq} is built from
\begin{agda}
⟦∘⟧ : ⟦ δ ∘ σ ⟧Th a ≡ ⟦ δ ⟧Th (⟦ σ ⟧Th a)
\end{agda}
which we can only prove by induction: it is a propositional, not a
definitional, equality. In the case \ag{t-k ∘ t-k} of \ag{⟦∘⟧} itself, the
element $v$ is transported along two \emph{different} paths, one coming from
\ag{∘-k q e} and one from composing the two stored equations, and we have to
show that the results agree. Those paths live in the semantic types
\ag{⟦ C ⟧T γ}, which are arbitrary types, so we need uniqueness of identity
proofs for them (the lemma \ag{pair-irr}, equivalent to UIP). Weakening causes
the same problem: \ag{⟦ wkTy A p ⟧T (γ , X)} is related to \ag{⟦ A ⟧T γ} only
by a map whose naturality (\ag{unwk-nat}) holds propositionally.

Note what the problem is \emph{not}. The recursive equations of the syntax,
such as \ag{δ ∘ tε = δ}, hold definitionally in the syntax, and the
interpretation simply follows the reduced terms. Nothing requires
\[
\llbracket \delta\circ\sigma\rrbracket = \llbracket\delta\rrbracket\circ\llbracket\sigma\rrbracket
\qquad\text{and}\qquad
\llbracket \mathrm{wk}\,A\rrbracket(\gamma,X) = \llbracket A\rrbracket\,\gamma
\]
to hold definitionally in the semantics: \ag{⟦∘⟧} and \ag{unwk-nat} are paths,
and that is enough to define the interpretation.\footnote{An earlier draft
of this note (in the repository) claimed that the interpretation requires these equations to hold
definitionally. This is incorrect. Definitional equations would be sufficient
(and for closed terms they do hold: for every numeral $n$,
$\llbracket\mathrm{SST}\,n\rrbracket$ computes to the expected iterated
$\Sigma$-type), but they are not necessary. What is needed is only that the
propositional equations are coherent.} The problem is their
\emph{coherence}: in the naive interpretation above a kept entry transports
its value along the image of a stored equation, and composing keeps produces
two transports that must agree. With arbitrary semantic types this is exactly
UIP.

In Agda the interpretation \agfile{Sem.agda} is K-free except for one module
parameter, \ag{pair-irr}, which is proved with K in \agfile{SemK.agda}. So this
naive interpretation does not give a semantic encoding of SST in HoTT.
Whether the coherences can instead be proved, at least when the semantic
types are truncated, is the subject of a separate note on a groupoid
semantics.

\begin{remark}
In two-level type theory the equations can be taken as \emph{outer}
(strict) equalities, which satisfy UIP. Read this way, the K-based
interpretation is essentially the known construction of semi-simplicial types
in 2LTT~\cite{ack-strict,acks-2ltt}. The open problem is precisely to get
rid of the outer level.
\end{remark}

\section{An alternative: composition as a constructor}

The obstruction comes from composition being a defined function whose
interpretation is only propositionally functorial. We could instead make
weakening and composition \emph{constructors}, and turn their recursive
equations into path constructors: a higher inductive-inductive type (file
\agfile{HIIT.agda}, a variant of Kovács' optimised syntax, see~\cite[Appendix~A]{rosain}).
\begin{itemize}
\item The semantic problem disappears. In the standard model composition is
  function composition, and every path constructor is interpreted by \ag{refl},
  without K.
\item The problem moves into the syntax. The keep rule now has to be stated
  as a path \emph{over} associativity,
\begin{agda}
t-k∘t-k : subst (λ d → Thin (ext A x d) _) ass (t-k δ ∘ t-k σ) ≡ t-k (δ ∘ σ)
\end{agda}
  so \ag{ass} is a path constructor, and it overlaps with the computation
  rules. For example, on \ag{(ρ ∘ δ) ∘ tε} we can go via \ag{ass} or via
  \ag{∘tε}; the two paths form a loop that nothing fills. These critical pairs
  would need 2-dimensional path constructors (the pentagon), which create
  critical pairs of their own, and so on.
\end{itemize}
The interpretation of the HIIT into the inductive-recursive syntax gives a map
HIIT $\to$ IR (file \agfile{HIITtoIR.agda}). The hope is that this map is an
equivalence (\agfile{HIITConj.agda}): then the HIIT is a set, and the IR syntax
inherits the K-free standard model. We have not carried this out. A sketch
suggests that, as stated, it fails under univalence: in a model where
\ag{Thin A B} carries an extra circle component and \ag{ass} is interpreted
with a twist by \ag{loop}, the loop above is sent to \ag{loop}, which is not
\ag{refl}.

\section{Fording the HIIT}\label{sec:fhiit}

The critical pairs of the HIIT come from \ag{ass}, and \ag{ass} is only there
because the keep rule has to say what the index of \ag{t-k δ ∘ t-k σ} is. We
can avoid this with the same trick as in Section~\ref{sec:ir}: Ford the HIIT. Every
index that would mention composition or weakening is replaced by a stored
path, and the rules that would have to compute such a path are stated for
\emph{every} path instead (file \agfile{experiments/forded-hiit/FHIIT2.agda};
the fragment without variables and weakening is in
\agfile{experiments/forded-hiit/FHIIT.agda}).
\begin{agda}
-- point constructors: composition and weakening are constructors
_∘_    : Thin B C → Thin A B → Thin A C
wk     : (E : Ty Γ) → Ty Γ → Ty (Γ ▷ E)
wkThin : (E : Ty Γ) → Thin A A' → Thin (wk E A) (wk E A')
t-k    : (σ : Thin A B) → δ ∘ σ ≡ δ' → Thin (ext A x δ') (ext B x δ)
zero   : wk A A ≡ C → Var (Γ ▷ A) C
suc    : Var Γ A → wk B A ≡ C → Var (Γ ▷ B) C

-- conditional path constructors: for every stored path on the right
t-k∘t-k : (r : δC ∘ (δ ∘ σ) ≡ δA) → t-k δ q ∘ t-k σ e ≡ t-k (δ ∘ σ) r
wkext   : (p : wk E B ≡ wk E B)
        → wk E (ext A x δ) ≡ ext (wk E A) (suc x p) (wkThin E δ)
wkt-k   : (e' : wkThin E δ ∘ wkThin E σ ≡ wkThin E δ') → ...
\end{agda}
together with the unconditional computation rules \ag{∘tε}, \ag{∘t-d},
\ag{t-d∘t-k}, \ag{wkε}, \ag{wkt-d}, \ag{wktε}. There is no \ag{ass}. The
right-hand side of \ag{t-k∘t-k} would need the path
\ag{ass ∙ cong (_∘ σ) q ∙ e}; quantifying over all \ag{r} means we never have
to name it. Similarly, \ag{wkext} is stated for every loop \ag{p}, not only
for \ag{refl}.

\paragraph{The syntax is a set.}
This version is a set, and this is formalised, without K and without
postulates:
\begin{agda}
isSetCon  : isSet Con
isSetTy   : isSet (Ty Γ)
isSetVar  : isSet (Var Γ A)
isSetThin : isSet (Thin A B)
\end{agda}
The proof is by normalisation. The normal forms are the inductive-recursive
syntax with codes of Section~\ref{sec:codes}, with \ag{zero} and \ag{suc} Forded as well (file \agfile{experiments/forded-hiit/NF2.agda}),
which is a set. There are maps
\ag{nf} from the HIIT to normal forms and \ag{quote} back, and
\ag{quote (nf t) ≡ t} is proved by induction on the HIIT. So every sort is a
retract of a set. Most of the work is in the path constructors, where one
has to fill a square, or for the weakening rules a cube. Two things make these
fillable.
\begin{itemize}
\item For a stored path we only ever have to compare two stored paths of the
  normal form, and those live in a set. So no coherence of the HIIT itself is
  needed.
\item The conditional rules give the needed freedom. In \ag{t-k∘t-k} the two
  sides of what would be a critical pair are two keeps that differ only in
  their stored paths, and the rule identifies them whatever the paths are.
  For \ag{wkext} this is essential: with \ag{refl} fixed in the weakened
  variable, the cube for \ag{wkext} has a boundary that can only be filled by
  a 2-cell in the HIIT, which is not available. With an arbitrary loop
  \ag{p}, the stored loop can be chosen so that every box closes.
\end{itemize}
(The proof uses \ag{TERMINATING} pragmas: the functions recurse through
path arguments of constructors and on the endpoints of path constructors,
which is structural but is not recognised by Agda's termination checker.)

\paragraph{Set-ness does not restrict the semantics.}
The conditional rules identify all stored paths, so they amount to UIP for
the \emph{syntax}. This is not an extra assumption: by encode-decode
(the codes of Section~\ref{sec:codes}, via the normal forms) the syntax is a
set anyway. Crucially, UIP for the syntax does not restrict the semantics. An
interpretation can first normalise, and interpret a stored path not by its
image but by the canonical semantic path computed from the normal forms, as
for the inductive-recursive syntax. Then the conditional rules are respected
automatically, since they become equations between normal forms, which form a
set.

\paragraph{The naive standard model needs UIP.}
The trouble is the naive interpretation, which interprets the HIIT as an
algebra: composition by composition of functions, and a keep by transport
along the interpretation of its stored path, as in
Section~\ref{sec:interp}:
\begin{agda}
⟦ t-k σ e ⟧ (a , v) = ⟦ σ ⟧ a , subst ⟦ x ⟧ (happly ⟦ e ⟧ a)⁻¹ v
\end{agda}
Composition is now interpreted as function composition, so this is fine. But
\ag{t-k∘t-k r} demands, for every path \ag{r}, that transporting $v$ first
along $\llbracket e\rrbracket$ and then along $\llbracket q\rrbracket$ gives the
same result as transporting along $\llbracket r\rrbracket$. The method of
the recursor for \ag{t-k∘t-k} receives the interpretation of \ag{r} as an
\emph{arbitrary} path between the semantic functions
$\llbracket A\rrbracket\,\gamma\to\llbracket D\rrbracket\,\gamma$, not only the
canonical one. So take it to be the canonical composite followed by a loop
$\ell$ (by function extensionality, a loop at each point). Then transport along $\ell$ must be the identity: in the semantic
telescope $\llbracket D\rrbracket\,\gamma$, \emph{every loop acts trivially on
the family} $\llbracket x\rrbracket\,\gamma$.

This is false. The context is arbitrary: $X_0$ is any type and $X_1$ any family
over $X_0\times X_0$. Take $X_0 = S^1$ and $X_1\,y\,z = (y \equiv z)$, and let
$\ell$ be \ag{loop} in the first point. Transport along $\ell$ in $X_1$ sends
$u : \mathsf{base}\equiv\mathsf{base}$ to $\mathsf{loop}^{-1}\cdot u$, which is not
$u$. The weakening rule fails even more directly. \ag{⟦ suc x p ⟧} transports
along $\llbracket p\rrbracket$, a loop in a family of types, that is, a family of
automorphisms of the telescope. Under univalence these exist, for instance the
swap of the two points of an edge telescope, and they do not act trivially. So
the conditional rules hold in the naive model only if all relevant identity
types are trivial: we are back to UIP, now for the semantic types. They hold
in the model of sets, where $X_0$ and the $X_k$ are sets. That model gives
semi-simplicial \emph{sets}, which we could define anyway.

\paragraph{The trade-off.}
Without the conditions the naive semantics would be fine, but then
\ag{t-k∘t-k} has to name its right-hand side, which brings back \ag{ass} and
the critical pairs. With the conditions the syntax is a set, but the naive
semantics needs UIP. An interpretation through normal forms avoids UIP in the
syntax, but then the coherence of the semantic paths has to be proved in the
semantics, as for the inductive-recursive syntax. Either way, the coherences
of composition have to be supplied somewhere, and stating them is the
original problem.

\section{Single substitutions and coherent syntax}\label{sec:ssc}

Rosain's report~\cite{rosain}, supervised by Kaposi, attacks the same problem
from a different side. It follows Kaposi's idea of a \emph{coherent} type
theory, one in which there is only one obvious way to prove two terms equal.
\begin{itemize}
\item The syntax is a HIT, not a QIT. Instead of parallel substitutions it uses
  the single substitution calculus of Kaposi and Xie~\cite{kaposi-xie}:
  weakening \ag{p}, lifting \ag{γ⁺} and instantiation \ag{⟨x⟩}, together with
  their computation rules.
\item In a dependent theory the computation rules for variables only typecheck
  up to equations between types, for instance
  \ag{vz[γ⁺] = vz} needs \ag{A[p][γ⁺] = A[γ][p]}. These equations are not
  computed but \emph{assumed}: the rule is stated for every path
  \ag{e : A[p][γ⁺] ≡ A[γ][p]}. These are conditional equations, exactly like
  our \ag{t-k∘t-k} and \ag{wkext}.
\item The plan is then to show that the syntax is a set by normalisation, that
  is, by pushing substitutions down to variables.
\end{itemize}
For the simply typed $\lambda$-calculus this works, and it is formalised:
types do not depend on terms, so there are no conditions. For the
variable-only dependent theory needed for semi-simplicial types it does not
yet work. To show that normal forms are stable under substitution one needs
the type equations, such as \ag{A[p][⟨y⟩] = A} and
\ag{A[p][γ⁺] = A[γ][p]}, generalised to substitutions lifted over a telescope.
The induction then asks for hypotheses about the type of a variable under
arbitrarily many lifted weakenings, which are too strong to be established
elsewhere. Generalising the type, or using iterated substitutions, leads to a
circular dependency. The report leaves the question open.

We think that this approach meets the same two problems as ours.

\paragraph{In the syntax: exchange instead of associativity.}
Single substitutions have no general composition, and so no \ag{ass}. We tried
the analogous idea for thinnings: a thinning is a word of single steps, a
face \ag{p} lifted under some keeps (file
\agfile{experiments/ssc/SSC.agda}). The
standard model is then all \ag{refl}, without K. But the
\emph{exchange law} $d_{j+1}d_i = d_id_j$, which is the shape of
\ag{A[p][γ⁺] = A[γ][p]}, is needed already at level~2 of SST, and its lifted
versions at level~3. Exchanges have critical triples. Three faces can be
reordered in two ways, forming a hexagon that needs a 2-cell. Four faces need
3-cells, and so on through the higher Bruhat orders.\footnote{Reordering
a word of $n$ steps by exchanging adjacent pairs is governed by the weak
Bruhat order on permutations of $n$ elements. The ways of rewriting with
exchanges are related by 2-cells, one hexagon for each triple, as in the
braid (Yang--Baxter) relation. These 2-cells are in turn related by 3-cells,
one for each quadruple, as in Zamolodchikov's tetrahedron equation, and so on.
Manin and Schechtman organised this hierarchy as the \emph{higher Bruhat
orders} $B(n,k)$~\cite{manin-schechtman}. Our exchange law for face maps is
not literally a braid relation, but its critical pairs, triples, \dots\ have
the same combinatorial shape.} The generalised
equations in Rosain's report, which need induction hypotheses under
arbitrarily many lifts, look to us like the syntactic shadow of this tower.
\ag{ass} has been traded for exchange, but there is still a tower of
coherences for composing non-invertible maps.

\paragraph{In the semantics: the naive model needs UIP.}
Suppose the normalisation argument went through and the coherent syntax were
a set. Its conditional equations would still block the naive standard model,
for the reason given in Section~\ref{sec:fhiit}. In the standard model
\ag{A[p][γ⁺]} and \ag{A[γ][p]} are both families over the same context, and
they are equal by \ag{refl}. The method for \ag{vz[γ⁺] =e vz} must work for
every semantic path $e$ between them, that is, for every loop in a type of
families. So transport along an arbitrary such loop must fix the generic
variable. Under univalence there are non-trivial loops, for instance an
automorphism of a two-point telescope swapping the points. The equation forces
them to act trivially, which brings back UIP. The conditional formulation is
what makes ``one obvious way'' true in the syntax, and the naive standard
model does not have this property. As for the Forded HIIT, this does not
exclude an interpretation through normal forms, but that moves the
coherences into the semantics.

So the single substitution calculus moves the problem around in the same way
as the Forded HIIT. Stating the equations conditionally makes the syntax a
set, or is meant to, but rules out the naive model in the universe. Stating them
for the canonical paths keeps the model but requires the coherences of
composition, here in the form of exchange.

\section{The obstruction is always there}

The attempts can be summarised as follows.
\begin{center}
\begin{tabular}{lll}
\toprule
presentation & syntax & naive model\\
\midrule
IR, index by $\delta\circ\sigma$ & transports block computation & ---\\
IR, Forded (\agdir{withK}) & needs UIP for \ag{Thin} & needs UIP\\
IR, codes (\agfile{TT.agda}) & a set, K-free & needs UIP\\
HIIT (\agfile{HIIT.agda}) & critical pairs (\ag{ass}) & all \ag{refl}\\
HIIT, Forded, conditional (\S\ref{sec:fhiit}) & a set, K-free & needs UIP\\
single steps (\agfile{experiments/ssc/SSC.agda}) & critical triples & all \ag{refl}\\
SSC, conditional (\S\ref{sec:ssc}) & a set? (open) & needs UIP\\
\bottomrule
\end{tabular}
\end{center}
Each version either puts the coherence of composition into the syntax, as
critical pairs, or into the semantics, where the naive interpretation needs
UIP to supply it. Whether the semantic coherences can be proved without UIP
when the semantic types are truncated is the subject of a separate note. What is never removed is the
coherence of composing \emph{non-invertible} maps, here thinnings. For
invertible structure, identity types and J supply all coherence for free (for
instance, coherently associative groups are exactly loop spaces). For one
non-invertible level nothing analogous is known in book HoTT. The same
obstruction appears in a minimal form for coherently associative semigroups
($A_\infty$ without units), which suggests that a solution would have to add
a genuinely directed principle, as two-level type
theory~\cite{acks-2ltt}, simplicial type theory~\cite{riehl-shulman} or
displayed type theory~\cite{dtt} do.

It is not clear, however, that displayed type theory resolves the problem.
Its construction of semi-simplicial types proceeds by d\'ecalage, and this
needs the display of the universe to consist of families,
$\mathcal U^{\mathsf d}\,A \simeq (A \to \mathcal U)$ (relativity). Sarah
Reboullet's thesis~\cite{reboullet} shows that in the nominal model of
internal parametricity this fails for the universe: bridges in $\mathcal U$
are not the same as families. So it is open whether there is a model of
displayed type theory with a universe for which relativity holds, and with it
whether displayed type theory yields semi-simplicial types in a homotopical
setting.

\section*{Use of AI tools}
This note and the accompanying Agda development were produced in
collaboration with Claude (Anthropic), used through Claude Code. Claude
drafted text and Agda code, carried out experiments in Agda, and searched the
literature, under the author's direction. The author has checked the content
and is responsible for it. Claims described as sketches have not been formally
verified.

\end{document}